\documentclass[aip,reprint]{revtex4-1}
\usepackage{mathtools}
\draft 

\begin{document}
	
	
	\title{Quantum-enhanced stimulated emission microscopy} 
	
	
	
	\author{Gil Triginer Garces}
	\affiliation{Department of Physics, University of Oxford, Clarendon Laboratory, Parks Road, Oxford OX1 3PU, United Kingdom}
	
	\author{Helen M. Chrzanowski}
	\affiliation{Department of Physics, University of Oxford, Clarendon Laboratory, Parks Road, Oxford OX1 3PU, United Kingdom}
	\affiliation{Humboldt University of Berlin, Unter den Linden 6, 10099 Berlin, Germany}
	
	\author{Shakib Daryanoosh}
	\affiliation{Department of Physics, University of Oxford, Clarendon Laboratory, Parks Road, Oxford OX1 3PU, United Kingdom}
	
	\author{Valerian Thiel}
	\affiliation{Department of Physics, University of Oxford, Clarendon Laboratory, Parks Road, Oxford OX1 3PU, United Kingdom}
	\affiliation{University of Oregon, 120 Willamette Hall, Eugene, OR 97403 USA}
	
	\author{Anna L. Marchant}
	\affiliation{Department of Physics, University of Oxford, Clarendon Laboratory, Parks Road, Oxford OX1 3PU, United Kingdom}
	
	\author{Raj B. Patel}
	\affiliation{Department of Physics, University of Oxford, Clarendon Laboratory, Parks Road, Oxford OX1 3PU, United Kingdom}
	
	\author{Peter C. Humphreys}
	\affiliation{Department of Physics, University of Oxford, Clarendon Laboratory, Parks Road, Oxford OX1 3PU, United Kingdom}
	
	\author{Animesh Datta}
	\affiliation{University of Warwick, Coventry CV4 7AL, United Kingdom}

	\author{Ian A. Walmsley}
	\affiliation{Department of Physics, University of Oxford, Clarendon Laboratory, Parks Road, Oxford OX1 3PU, United Kingdom}
	\affiliation{Imperial College London, South Kensington, London SW7 2AZ, United Kingdom}

	
	\begin{abstract}
	Nonlinear optical microscopy techniques have emerged as a set of successful tools for biological imaging. Stimulated emission microscopy belongs to a small subset of pump-probe techniques which can image non-fluorescent samples without requiring fluorescent labelling. However, its sensitivity has been shown to be ultimately limited by the quantum fluctuations in the probe beam. We propose and experimentally implement sub-shot-noise limited stimulated emission microscopy by preparing the probe pulse in an intensity-squeezed state. This technique paves the way for imaging delicate biological samples that have no detectable fluorescence with  sensitivity beyond standard quantum fluctuations.
	\end{abstract}
	
	
	\maketitle 
	

	Fluorescence is a backbone of optical microscopy of biological systems, underlying the spectacular advances in super-resolution microscopy over the past two decades \cite{Hell94, Hess06, Rust06}. However, several important proteins such as haemoglobin and cytochromes cannot be detected by fluorescence because their spontaneous emission is dominated by rapid non-radiative decay. To overcome this challenge, many methods begin by labelling a sample of interest with fluorescent proteins. This allows for the imaging of the sample, although at the risk of interfering with its biochemical properties. To avoid this unwanted side effect, Min et al. \cite{MinXie09} developed stimulated emission microscopy (SEM), a nonlinear microscopy technique able to image chromophores with undetectable fluorescence. However, it was shown that the sensitivity of SEM is practically limited by the shot noise of the stimulation beam, which is reflected in the lowest detectable concentrations for a given light dose or acquisition time. The measurement sensitivity can be improved by increasing either or both of the latter, but at the cost of fast photobleaching and formation of chemical radicals \cite{Cremer16}.
	
	Quantum physics has opened a new paradigm for optical sensing by leveraging non-classical correlations in either the probe light, the interaction with the sample, or the detection stage. The achievements of quantum metrology include demonstrations of imaging with sensitivity surpassing the shot noise limit ~\cite{OnoTak13,SamGen17, MorPad19}, enhanced resolution beyond the diffraction limit using quantum correlations~\cite{TenOro19}, and tracking of motion, such as the movement of lipid granules diffusing through cells~\cite{Taylor12}.
	
	Here we propose to improve the sensitivity of SEM by reducing the fundamental shot noise in  the stimulation beam. In contrast to classical SEM, we prepare the stimulation beam in an intensity-squeezed coherent state of light \cite{Wal83}.  This optical state exhibits sub-Poissonian photon statistics, i.e. the variance of the photon flux is smaller than its mean. As a consequence of the reduced intensity fluctuations in the measurement signal, the small amplification induced by the sample can be resolved more precisely. We prepare an ultrafast intensity-squeezed probe,  suitable for this non-linear microscopy technique, via the deamplification of coherent seed pulses in a single-pass optical parametric amplifier (OPA). In addition, we perform shot-noise-limited direct detection of the intensity squeezed light, and observe reduced quantum fluctuations in the SEM signal.  
	
	Many chromophores have very short-lived excited states with much faster non-radiative decay rates than their spontaneous emission rates. As a result, their feeble fluorescence is overwhelmed by background and detector dark counts. Nevertheless, an excited molecule can be stimulated down to the ground state by an incident light field with a frequency corresponding to the energy of a transition in the molecule, which results in the creation of a new coherent photon identical to those in the original incident field. The number of additional emitted photons per incident quantum of light is proportional to the number of excited molecules in the sample. SEM uses this optical amplification phenomenon as a contrast mechanism for imaging chromophores with undetectable fluorescence \cite{MinXie09}. 
	
\begin{figure}[t]
		\centering
		\includegraphics[width=.92\linewidth]{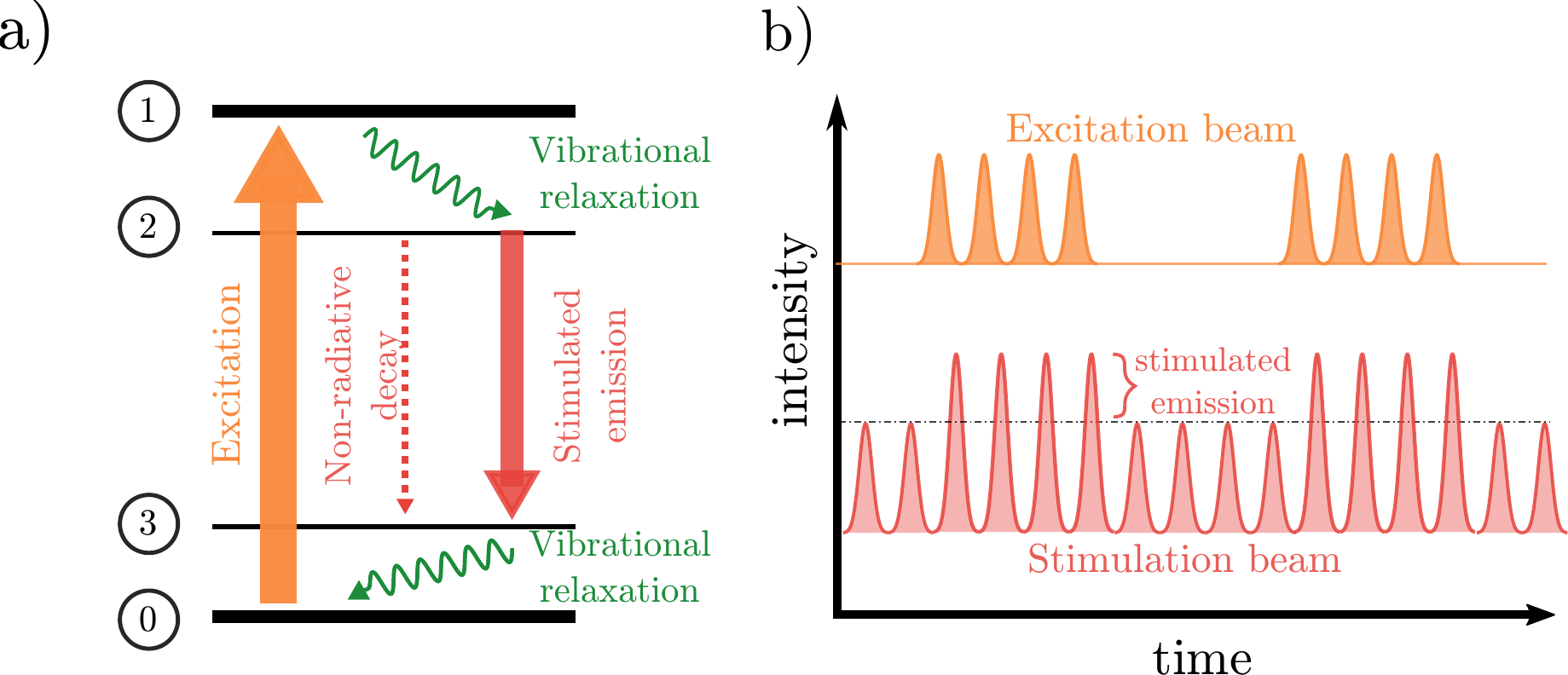}
		\caption{\label{fig:sem} a) Energy level diagram for stimulated emission-based pump-probe microscopy.
			b) Amplification of the stimulation beam synchronised with the modulated excitation beam.}
	\end{figure}

	In SEM, an excitation beam with intensity $I_{\rm E}$ , incident on a collection of chromophores, promotes them to an excited state (level $1$ in Fig.~\ref{fig:sem}a). Subsequently, a stimulation beam with a longer wavelength probes the sample, resulting in emission of light in the mode of the impinging stimulation beam. As a consequence, optical amplification takes place (Fig.~\ref{fig:sem}b). Assuming that the intensity of the stimulation beam is $I_{\rm S}$, the intensity of the generated stimulated emission, $\delta I_{\rm S}$, is\cite{MinXie09} $\delta I_{\rm S} = I_{\rm S} N_2 \sigma_{\rm s}/A \propto I_{\rm S} I_{\rm E} N_0 \sigma_{\rm s}\sigma_{\rm a}/A^2$, where $N_k$ is the population in the $k$-th energy level (Fig. \ref{fig:sem}.a), $A$ represents the waist area of the beam, and $\sigma_{\rm a}$ and $\sigma_{\rm s}$ are the absorption ($0\to 1$ transition) and stimulation ($2\to 3$ transition) cross sections, respectively. Here, it is assumed that under a non-saturating regime of the dynamics, the population of the excited state satisfies $N_2 \propto N_0 I_{\rm E} \sigma_{\rm a}$.
	The above expression shows that, after the interaction, the intensity of the stimulation beam is amplified by a factor $G_{\rm SE} = (1 + \delta I_{\rm s}/I_{ \rm s})$, which in realistic experimental conditions of weak interaction strength is very close to unity.  The goal of SEM is to detect the small variations in $G_{\rm SE}$ as the beams are scanned across a sample, which provides an estimation of the number of molecules in the ground state, $N_0$. Given that the stimulated emission gain is proportional to the overlap of the excitation and stimulation fields, this method offers intrinsic three-dimensional sectioning at the focci of the beams, as well as a finer spatial resolution than absorption imaging. 
	
	However, due to the weakness of the SEM signal, its measurement is hindered by the low-frequency technical noise of the laser and electronics. In order to mitigate this technical noise, W. Min et al.~\cite{MinXie09} implemented a modulation of the excitation beam at a few MHz, such that the stimulation emission signal is moved to a sideband frequency beyond the technical noise in the system, and can be separated using microwave filters. In this regime, the predominant limitation to the measurement of $G_{\rm SE}$ is the wideband shot noise in the stimulation beam \cite{MinXie09}. 
	
	In a standard SEM experiment, the stimulation laser is well described as a train of ultrafast optical pulses in a coherent state. After traversing a sample together with a suitable excitation beam, and thereby picking up a time-dependent gain, these pulses are detected with a photodiode.  In order to distinguish the small periodic amplification from the background noise, the resulting photocurrent is measured at the modulation frequency of the gain, e.g. by means of a spectrum analyser (SA). We will now briefly introduce a theoretical model of this measurement\cite{supplemental}, which suggests that the fundamental sensitivity limit set by the shot noise (that is, the quantum intensity fluctuations associated with coherent states of light) can be overcome using squeezed light. 
	
	We model the optical gain in SEM as a phase-independent linear amplification mechanism and assume that amplification noise can be ignored in the weak gain regime\cite{supplemental}. We allow the intensity gain to evolve in time according to $G_{\rm SE}(t) = G_0 + (G_0 -1) \, \cos(\Omega_0 t)$,  where the gain parameter $G_0$ is close to unity and $\Omega_0$ is an RF modulation frequency. As a result, the power spectral density (PSD) measured in the SA contains a narrowband component at $\Omega_0$. The amplitude of this signal is proportional to $\mathcal{S}_{\rm out}(\Omega_0) = (G_0 -1)^2 P_0^2/16$, where $P_0$ is the average optical power in the probe beam before the amplification process. The dependence of $\mathcal{S}_{\rm out}(\Omega_0)$ on $G_0$ makes this signal a suitable proxy to estimate the sample density. 
	
	However, in addition to this narrowband signal, the measured PSD also contains noise of various origins. In the first place, the quantum fluctuations in the intensity of the probe field appear as a white (frequency-independent) noise contribution around the signal peak. According to a simple description of optical quantum noise, and neglecting SEM amplification for the moment, the noise spectral density of quantum origin is $\mathcal{N}_{\rm in}^q = \kappa P_0 F $.  This shows the characteristic linear proportionality of the noise to the optical power, $P_0$, scaled by a constant that accounts for the particularities of the experimental implementation,  $\kappa$. The constant $F$ is analogous to the Fano factor of a random variable: it takes the value $F=1$ for a coherent beam (Poissonian statistics), and $F<1$ or $F>1$ for light fields with sub-Poissonian and super-Poissonian photon statistics respectively. In the presence of the modulated SEM gain, the quantum noise spectral density is amplified to $\mathcal{N}_{\rm out}^q = \mathcal{N}_{\rm in}^q \ G_0$. In our model, we consider a second source of noise: low-frequency intensity fluctuations of classical origin \cite{Ivanov03}. In the absence of SEM amplification, the contribution of classical technical noise, which we denote as $\mathcal{N}_{\rm in}^t(\Omega)$, is assumed to be non-zero only at frequencies well below $\Omega_0$. However, in the presence of gain, a fraction of this noise is modulated up in frequency, yielding a noise spectral density contribution $\mathcal{N}_{\rm out}^t(\Omega) = (G_0 -1)^2 \ \mathcal{N}_{\rm in}^t(\Omega + \Omega_0)/4$. 
	
	The signal to noise ratio (SNR) of the SEM measurement, taken as the ratio between the signal and noise components of the measured PDF at the modulation frequency $\Omega_0$, is then\cite{supplemental}:
	\begin{equation}\label{SNR_main_text}
	{\rm SNR} = \frac{\mathcal{S}_{\rm out}(\Omega_0)}{\mathcal{N}_{\rm out}^q(\Omega_0) + \mathcal{N}_{\rm out}^t(\Omega_0)} =\frac{(G_0-1)^2 P_0 /\kappa}{16 F G_0 + (G_0-1)^2 \rho_t}
	\end{equation} 
	where we have defined $\rho_t = \mathcal{N}_{\rm in}^t(0) / (P_0 \kappa)$, which corresponds to the ratio between the spectral density of the low-frequency technical noise and the quantum shot noise in a coherent state ($F=1$). It can be seen that, by increasing the power of the stimulation beam, $P_0$, the SNR improves, but only at the risk of damaging the imaged sample. To avoid this, we instead propose to probe the sample with an intensity-squeezed coherent state of light \cite{Sch17}. Such a state exhibits sub-Poissonian intensity statistics, i.e. $F<1$, yielding a higher SNR in the estimation of $G_{\rm SE}$ without increasing the intensity of the probe. While a lower $F$ inmediately implies a greater SNR, it is worth noting that the enhancement will be largest when the technical noise is small compared to the quantum fluctuations. This condition is met in the low-gain regime that we are interested in, since technical noise contribution to the SNR is quadratic on $G_0$, while the quantum shot noise contribution is linear.
	
	A squeezed coherent state can be generated via the optical parametric amplification (OPA) of a \textit{seed} coherent state~\cite{Wal83}. After an interaction with a suitable \textit{pump} field, mediated by a nonlinear medium, the seed experiences a phase-sensitive amplification/deamplification which renders its fluctuations quantum-correlated. Using a simple model of OPA\cite{supplemental}: considering a pump and seed fields with average powers $P_{\rm p}$ and $P_{\rm s}^{\rm in}$ respectively, the  maximum/minimum powers of the amplified/deamplified seed, obtained for the corresponding extreme phase offsets between the seed and the pump, are: 
	\begin{equation} \label{classic:amp}
	P_{\rm s}^{\rm out} =   \Big(1-\chi + \chi \exp\big(\pm \beta \sqrt{P_{p}}\big)\Big) \, \eta \, P_{\rm s}^{\rm in},
	\end{equation}
	where $\beta$ is an effective nonlinearity, $\chi \in [0,1]$ is a coefficient that characterizes the mode overlap between the two interacting fields, and $\eta \in [0,1]$ is an efficiency that compounds the propagation loss and the detection efficiency. 
	 
	In this situation, our model predicts the following maximum/minimum intensity Fano factor for the amplified/deamplified seed\cite{supplemental}:
	\begin{equation}\label{multimode_Fano}
	F \approx 1 - \eta + \eta \, \frac{1 - \chi + \chi \, \exp(\pm 4 \,\sqrt{P_{\rm p}} \, \beta)}{1 - \chi + \chi \, \exp(\pm 2 \, \sqrt{P_{\rm p}} \,  \beta)}.
	\end{equation}
	This simple model offers an interesting insight: while single-pass OPA allows us to achieve sub-Poissonian photon statistics in the seed field ($F<1$), in the presence of modal mismatch ($\chi < 1$) the degree of intensity squeezing measured by direct detection does not grow monotonically with the pump power. Instead, the dependence of the sub-Poissonian Fano factor on the pump power shows a turning point, beyond which it asymptotically approaches $F \rightarrow 1$.   
	
	The model presented here is meant to provide some physical intuition without being fully rigurous. A full account of parametric deamplification of a pulsed coherent beam must incorporate a number of phenomena that have not been considered here (e.g. group velocity mismatch between the pump and seed pulses, beam divergence, spatial walk-off or spatio-temporal coupling in the parametric gain). A detailed account of the theoretical description and design of a source of pulsed sub-Poissonian light with a finite transverse profile will be provided in a future publication \cite{TraWal20}.
	
\begin{figure}[t]
		\centering\includegraphics[width=.75\linewidth]{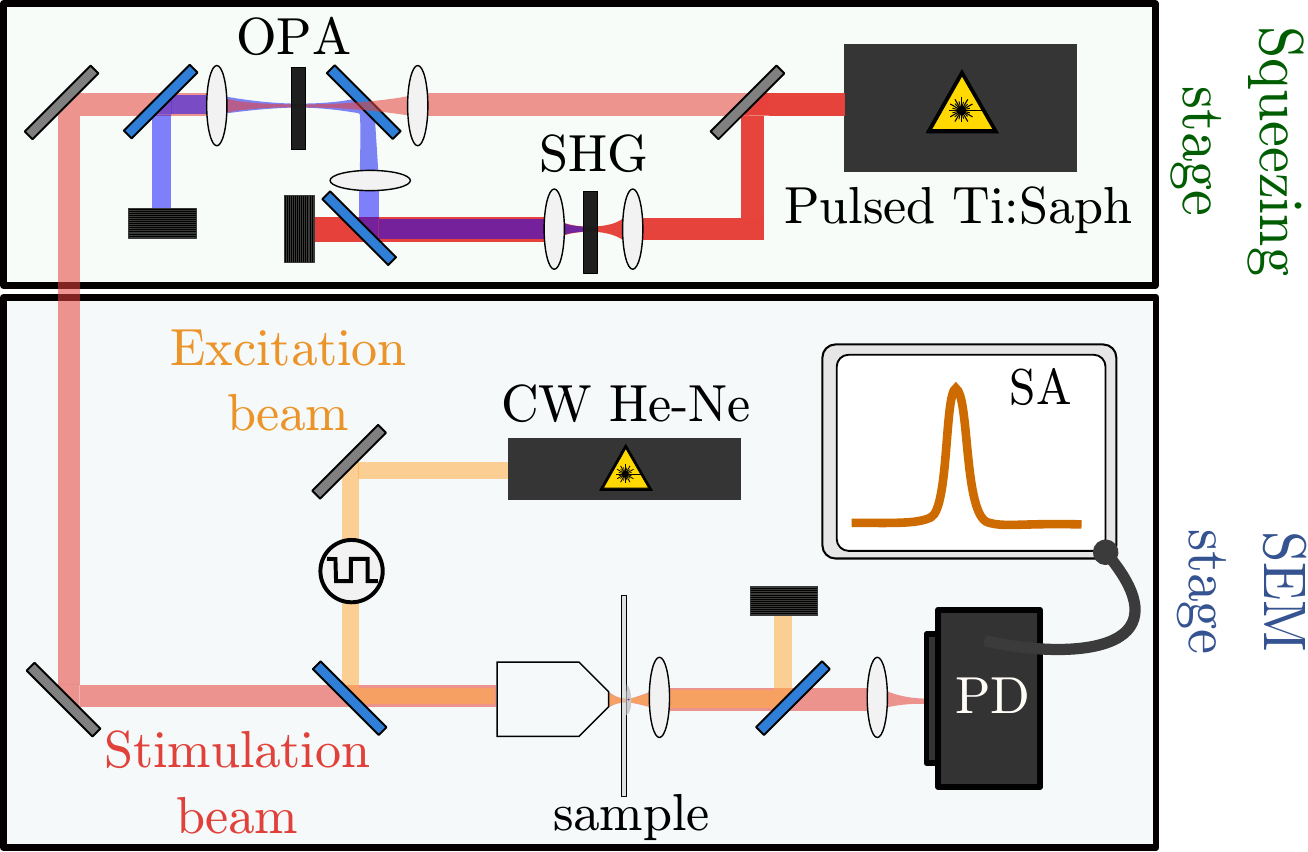}
		\caption{\label{experimental_setup_figure}
			 Experimental setup for intensity squeezed light generation (top) and SEM stage (bottom). PD: photodiode; SA: spectrum analyzer.
		 }
	\end{figure}

	Our experimental setup can be divided into two parts: an intensity-squeezing setup and an imaging apparatus (see Fig. \ref{experimental_setup_figure}). Our sub-Poissonian light source is implemented in a single-pass degenerate OPA process, where the pump and seed fields are provided by a femtosecond Ti:Sapphire laser with a repetition rate of $80$ MHz, central wavelength of  $\lambda_{\rm S}=820$ nm, pulse duration of $180$ fs, and maximum power of $2.8$ W. The fundamental Ti:Saph is spatially filtered and split into a strong and weak beams, where the first is frequency-doubled in a second-harmonic generation (SHG) setup ($0.4$mm-long type-I BiBO) and the latter will act as the \textit{seed} beam in the OPA. After temporally and spatially overlapping them, the high-intensity pump and low-power seed interact inside a $0.3$ mm-long, degenerate, type-I BBO optical parametric amplifier, after which the pump is filtered out using a low-loss  dichroic mirror. This leads to amplification/deamplification of the seed depending on its phase relative to the pump. We set this phase to yield maximum deamplification and squeezing using a piezo-controlled mirror in a feedback control loop configuration. To implement the feedback, the phase of the signal field is very weakly dithered, yielding an error signal proportional to the phase difference between the pump and seed fields when mixed down at the sideband frequency. We verified that this phase modulation of the seed did not introduce a significant mixing of the noise in its squeezed and antisqueezed quadratures.
	
	In a first characterization step, this signal can be sent directly to a photodetector in order to evaluate the degree of intensity squeezing. We employ a custom-made photodetector which achieves shot-noise limited direct detection of the deamplified seed in the RF sideband where we place the SEM signal ($\sim$ 4 MHz). Our detector employs a Hamamatsu S5971 silicon PIN photodiode with 85\% quantum efficiency at $820$nm, and and has a bandwidth of $100$MHz when reverse-biased at $30$V.  The photocurrent is measured in the frequency domain using a high dynamic range signal analyser. 
	
	The SEM imaging stage involves two fields, the excitation beam and the sub-Poissonian stimulation beam, which interact inside the imaged sample. The stimulation beam is prepared in the intensity-squeezing stage described above. The excitation beam is provided by a continuous-wave (CW) He-Ne laser with $\lambda_{\rm E} = 632.8$ nm. This field is intensity-modulated with an acousto-optical modulator in order to move the stimulated emission gain to an RF sideband at 4 MHz.  Note that, ideally, one would excite the sample with ultrafast pulses to maximize the field intensity whilst minimizing average power, allow for temporal suppression of additional processes, and avoid deleterious nonlinear effects \cite{MinXie09}. For the proof-of-principle experiment conducted here, we assume that these considerations can be neglected and use a modulated CW laser. The imaged sample is a diluted dye, Brilliant Blue FCF (an organic molecular compound), with a strong absorption band near $\lambda_{\rm E}$, which is well separated from the spectrum of the stimulation beam. The excitation and stimulation beams are combined on a dichroic mirror, and focused into the sample using an objective, achieving a beam waist of approximately 50 $\mu$m. After the interaction, the excitation beam is filtered out from the stimulation beam by means of two spectral filters. The intensity in the stimulation beam is measured with the photodetector + RF spectrum analyzer arrangement described above, where a spectral peak at the frequency of the acousto-optic modulator confirms the existence of stimulated emission. The spectrally-flat noise background that surrounds this peak corresponds to optical intensity fluctuations. 
	
\begin{figure}[t]
		\centering\includegraphics[width=9cm]{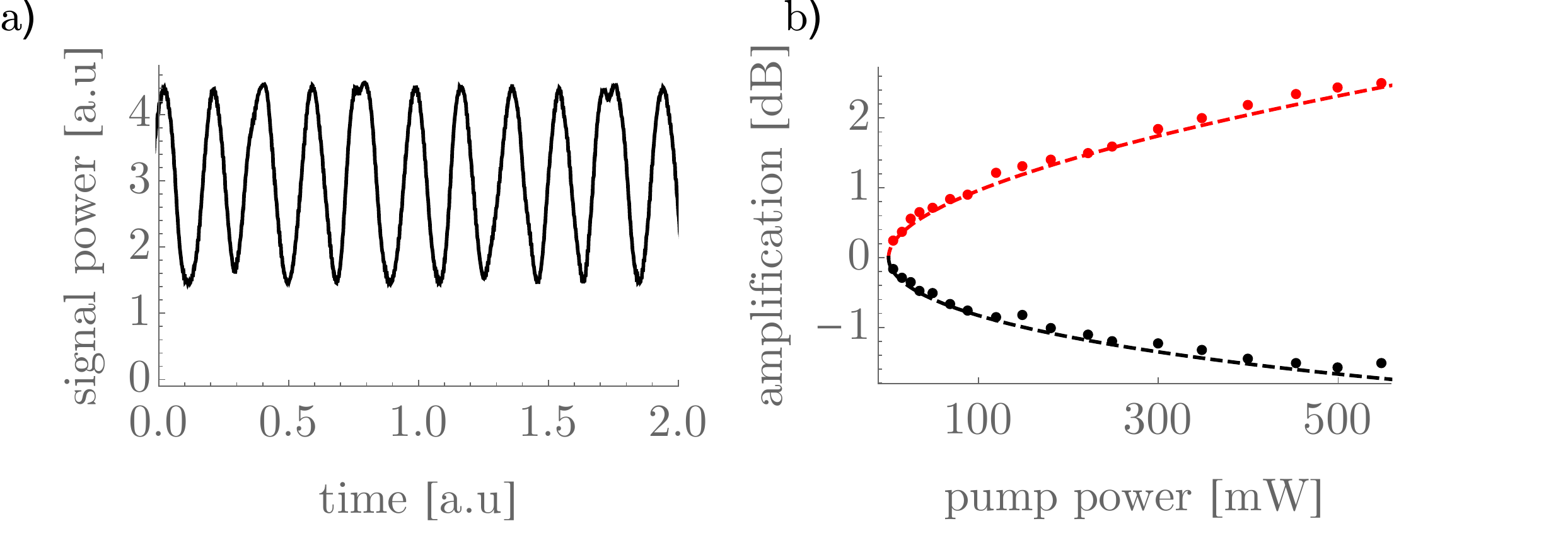}
		\caption{\label{fig:amp} a) Amplification/deamplification of the signal as its phase is linearly scanned in time. b) Maximum amplification (red dots) and deamplification (black dots) as a function of the pump power. The fitted curves (dashed lines) take into account the imperfect mode overlap between the pump and signal, Eq.~(\ref{classic:amp}). }
	\end{figure}

	Let us begin by reporting the performance of our sub-Poissonian light source in the absence of the SEM imaging stage. The degree of intensity squeezing achieved in a single-pass OPA is correlated with the amount of classical intensity deamplification\cite{supplemental}. Figure~\ref{fig:amp}(a) shows the average power of the output signal as the relative phase between the pump and seed is scanned. A fit of our measurements to the simple model presented in Eq. (\ref{classic:amp}) is shown in Fig.~\ref{fig:amp}(b), where the best fit is obtained for a mode overlap parameter $\chi = 0.58$.
	
\begin{figure}[t]
		\centering\includegraphics[width=.95\linewidth]{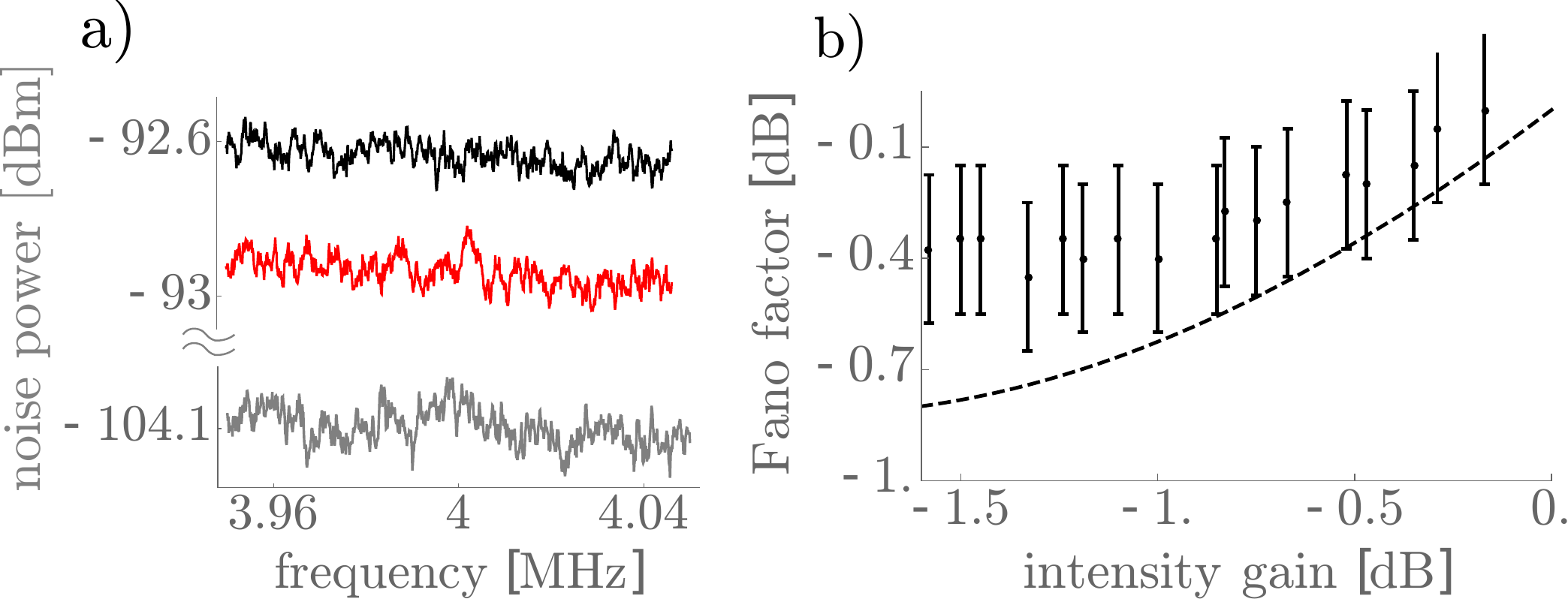}
		\caption{\label{fig:squeezing} a) Noise power of a classical (black) and deamplified seed (red) beams of matching mean intensity, and electronic noise of the SA (grey). The resolution bandwidth is $1$ kHz, and the video bandwidth is $10$Hz. b) Fano factor versus mean intensity deamplification for different pump powers. The dashed line is a fit to Eq.(\ref{multimode_Fano}). }
	\end{figure}

	The intensity fluctuations of the deamplified seed are expected to be smaller than those of a coherent field of the same average power. Figure~\ref{fig:squeezing}(a) shows the noise power of our deamplified seed in the frequency range of interest (red). This is compared with the noise power of the same field in the absence of parametric deamplification (black). The average power of the amplified and non-amplified fields are matched using a waveplate + polarizer power control stage. The electronic noise (grey) is almost an order of magnitude below the shot noise, and we have verified that, beyond an analysis frequency of 2 MHz, the shot noise exhibits the flat frequency dependence characteristic of quantum fluctuations. The gap between the noise power of the deamplified and non-deamplified seeds clearly signals a Fano factor $F<1$ of the former, proving its sub-Poissonian statistics. 
	
	The degree of intensity squeezing is expected to increase with pump power, similarly to the amount of classical intensity deamplification. In Fig.~\ref{fig:squeezing}(b) the Fano factor of the deamplified seed is plotted as a function of the classical intensity deamplification. In contrast to classical intensity deamplification, the amount of intensity squeezing approaches a plateau. The discrepancy between squeezing and classical deamplification can be attributed to propagation losses, inefficient detection, and imperfect mode overlap between the pump and seed beams (e.g. spatial or temporal). We determine the propagation efficiency to be $\eta_{\rm p} = 0.85$, and the detector efficiency to be $\eta_{\rm d} = 0.85$. Using the effective nonlinear parameter and mode overlap obtained in the previous classical deamplification fit, and adding the effect of propagation loss and detection inefficiency, we compute the expected Fano factor using Eq. (\ref{multimode_Fano}). While this calculation fits the experimental data for low values of the classical deamplification, it breaks down at higher gains. To understand this discrepancy, a more accurate model of OPA of a pulsed coherent beam is needed~\cite{TraWal20}.
	
\begin{figure}[t!]
		\centering
		\includegraphics[width=.82\linewidth]{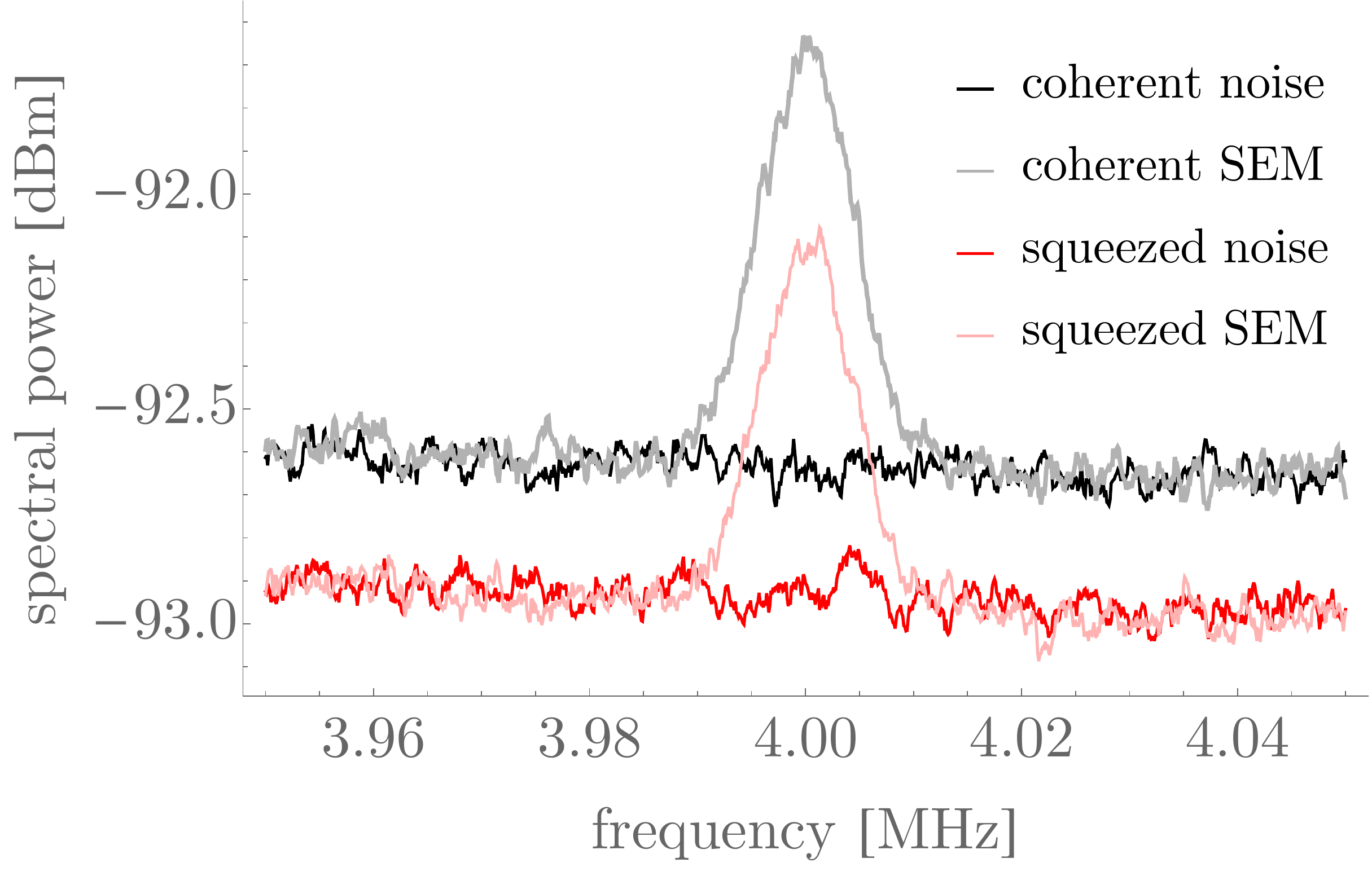}
		\caption{\label{fig:SNR} Stimulated emission microscopy signals for an intensity-squeezed probe (light red) and coherent beam of the same average power (grey). The shot-noise limit is shown in black and the noise power of the squeezed probe is in red.}
	\end{figure}

	We now proceed to evaluate the performance of our squeezing-enhanced SEM setup. First we quantify the degradation of squeezing due to the  non-ideal transmission efficiency ($75\%$) of the microscope objective and the dye. As can be seen in Fig.~\ref{fig:SNR},  a residual squeezing of $0.3$ dB is measured. Next we assess the impact of this quantum enhancement on the SEM measurement. The light-red and grey traces in Fig.~\ref{fig:SNR} correspond to the measured noise power for a squeezed and classical probes, in the presence of an excitation beam which allows stimulated emission to occur. The average power of the stimulation and excitation beams are set respectively to $2$ mW and $30 \, \mu$W to avoid excessive photobleaching~\cite{SonTan95}. The appearance of a spectral peak at the frequency of modulation of the excitation beam, and
	its absence when the stimulation beam is blocked, signals the presence of stimulated emission. As is evident, the shot noise is reduced, which indicates an improvement in the signal to noise ratio. This can be shown from Eq. \ref{SNR_main_text}, which gives the fraction improvement as
	\begin{equation}
	\Delta_{\rm SNR} = \frac{{\rm SNR}_{\rm squeezed}}{{\rm SNR}_{\rm coherent}}= \frac{1 + (G_0-1)^2 \rho_t / (16 G_0)}{F + (G_0-1)^2 \rho_t / (16 G_0)}
	\end{equation}
	This is always larger than unity for $F < 1$, no matter how much technical noise is present. The quantitative improvement of course depends on the latter, and in the present case, this can be estimated. We estimate the sideband stimulated emission gain $(G_0-1)$ to be smaller than $1\%$, on the grounds that a greater amplification would have been resolved by our photodetection setup in the absence of gain modulation. The value of $\rho_t$ is harder to gauge, and is not simple to measure directly.  Both the technical noise power and the input photon flux are positive numbers, and a reasonable commercial laser would have a ratio $\rho_t < 1$. Allowing for additional noise due to, for example, mode overlap fluctuations, sample motion and photobleaching would increase this, but it would not be excessive to put an upper bound of $\rho_t \sim 1$. This implies that the contribution of the technical noise to the SNR is minor. 
	Therefore, the observed reduction of the shot noise provides evidence that, according to the theoretical
	model of SEM that we have presented, the average SNR of the measurement was improved by at most 0.3 dB.
	
	In summary, we have proposed and implemented a quantum-enhanced nonlinear microscopy technique which employs stimulated emission to remove the requirement for labelling biological samples. We have identified that shot noise fluctuations in the stimulation beam currently pose a real limitation in the sensitivity of SEM, and have reduced these by preparing an ultrafast probe beam in an intensity-squeezed state. Using a single-pass degenerate OPA,  we have achieved direct observation of macroscopic light with a Fano factor of $-0.4$ dB
	. However, further noise reduction has been prevented by spatio-temporal mode mismatch and mode mixing in the OPA pump and seed pulses, as well as by imperfect detection efficiency. We have used this nonclassical source of light in conjunction with an SEM imaging setup, observing that sub-Poissonian statistics persist in the measurement signal. Following the theoretical model of SEM that we have developed, this reduction of the shot noise implies an improvement of the SNR of the measurement. 
	
	This project was supported by the Engineering and Physical Sciences Research Council (EPSRC) Quantum Technology Hub in Quantum Enhanced Imaging (QuantIC) EP/M01326X/1.  G.T thanks Merton College, Oxford, for its support. 

	We thank N. Treps and J. Francis-Jones for fruitful discussions. 
	
	The data that support the findings of this study are available from the corresponding author upon reasonable request.
	
	
	%
	%
	
	%
	

	
	\bibliography{references}
	
	\appendix 
	
	\section{Model of signal and noise in stimulated emission microscopy}
	\label{appendix_model_SEM}
	Let us treat the stimulated emission process as a phase-independent linear amplification mechanism, such that the annihilation operator that describes the probe field is transformed according to 
	
	\begin{equation} \label{phase_insensitive_gain}
	\hat a_{\rm out} = \sqrt{G_{\rm SE}} \, \hat a_{\rm in}+ \hat L^\dagger,
	\end{equation}  
	where for the moment we are omitting the temporal dependence of the fields and gain, and where the second term in the right-hand side is an operator that accounts for the fundamental noise added in a phase-insensitive amplification process \cite{Caves82, Caves12, ComCav16}.  In order to preserve the commutation relations of the output bosonic operators ($\hat{a}_{\rm out}$) if the input field ($\hat{a}_{\rm in}$) is uncorrelated with the amplifier noise ($\hat{L}$), the latter has to satisfy $[\hat{L}, \, \hat{L}^\dagger] = G_{\rm SE} -1$. In particular, a phase-insensitive amplifier can be modelled by taking the amplifier ancilla mode $\hat{L}$ to be in the vacuum state \cite{Caves12}. It can be seen that the photon flux at the output of the amplifier is then
	\begin{equation}
	\langle \hat{n}_{\rm out} \rangle = G_{\rm SE} \, \langle \hat{n}_{\rm in} \rangle + (G_{\rm SE} -1)
	\end{equation}
	where $\hat{n}_{\rm in \, (out)} = \hat{a}_{\rm in \, (out)}^\dagger \hat{a}_{\rm in \, (out)}$ are the input and output optical fluxes, and where the last term in the right-hand side corresponds to amplified spontaneous emission. The variance of the photon flux at the output of the amplifier is 
	\begin{equation} \label{eq:variance_SEM_noise}
	\begin{split}
	{\rm var}(\hat{n}_{\rm out}) &= G_{\rm SE}^2 \, {\rm var}(\hat{n}_{\rm in}) + G_{\rm SE} \, (G_{\rm SE} -1) \langle n_{\rm in} \rangle \\
	&+ G_{\rm SE} \, (G_{\rm SE} -1),
	\end{split}
	\end{equation}
	where the last term is negligible in experiments involving a small gain and an input with a macroscopic number of photons. In our experiment, the stimulated emission generated in the sample is so weak, that even the second term in the r.h.s. becomes negligible. Considering that the stimulated emission generated in the sample is too weak to be measured with a DC photodiode, we can make the conservative estimation that the stimulated emission gain is at most $G_{\rm SE} < 1 + 10^{-2}$. Therefore, the main source of amplification noise, represented by the second term in the r.h.s. of Equation (\ref{eq:variance_SEM_noise}), is at least two orders of magnitude below the shot noise, and thus does not affect our measurements.
	
	Let us then simplify our model by considering that the SEM interaction can be described by
	\begin{equation}
	\hat{n}_{\rm out}(t) \approx \hat{n}_{\rm in}(t) \, G_{\rm SE}(t)
	\end{equation}
	where the phase-insensitivity of the process becomes apparent, and where the noise introduced in the amplification is ignored. The photon flux before the interaction can be described as a train of pulses
	\begin{equation} \label{input_pulse_train}
	\hat{n}_{\rm in}(t) = \sum_n \phi(t - n T) + \delta \hat{\phi}(t)
	\end{equation}
	where $\phi(t)$ is the mean intensity envelope of a single pulse, whose temporal duration is in the order of the femtoseconds, and where the repetition rate of the laser, $1/T$, is in the order of the tens of MHz. The operator $\delta \hat{\phi}(t)$ describes the intensity fluctuations in the input probe, which we assume to be described by Gaussian noise with a zero mean, $\langle  \delta \hat{\phi}(t) \rangle = 0$. 
	
	At this point, it is useful to provide a description of the frequency-resolved photodiode current. We characterize this current using an RF spectrum analyser (SA) which provides an estimation of the power spectral density (PSD) of the photocurrent. In the limit where the integration time of the SA is long, and assuming a unit photodetection efficiency for the sake of clarity, the estimated PSD will be
	\begin{equation}
	{\rm PSD}_{\rm in (out)}(\Omega) \propto \langle \hat{N}^\dagger_{\rm in (out)}(\Omega) \hat{N}_{\rm in (out)}(\Omega) \rangle,
	\end{equation}
	where we have defined the input (output) frequency-resolved photon fluxes
	\begin{equation}\label{frequency_resolved_photon_flux}
	\hat{N}_{\rm in (out)}(\Omega) = \int {\rm d}t \, \exp(i \Omega t) \, \hat{n}_{\rm in (out)}(t).
	\end{equation}
	
	This PSD can be split into a \textit{signal} component with a non-stochastic origin, \begin{equation}\label{PSD_signal_component}
	\mathcal{S}_{\rm in(out)}(\Omega) = |\langle \hat{N}_{\rm in(out)}(\Omega)\rangle|^2,
	\end{equation}
	and a \textit{noise} component with a stochastic origin, 
	\begin{equation}\label{PSD_noise_component}
	\mathcal{N}_{\rm in(out)}(\Omega) = \langle  \hat{N}^\dagger_{\rm in(out)}(\Omega)\,\hat{N}_{\rm in(out)}(\Omega)\rangle - \mathcal{S}_{\rm in(out)}(\Omega).
	\end{equation} 
	
	Let us begin by evaluating these two terms for the pulsed input probe given in equation (\ref{input_pulse_train}). In the first place, the signal term corresponds to a frequency comb, given by
	\begin{equation}\label{input_signal}
	\mathcal{S}_{\rm in}(\Omega) = |\langle \hat{N}_{\rm in}(\Omega)\rangle|^2 = \Big|\Phi(\Omega) \sum_n \delta(\Omega - n/T)\Big|^2,
	\end{equation}
	where $\Phi(\Omega) = \int dt \, \exp(i\Omega t) \, \phi(t) $, such that the bandwidth of $\Phi(\Omega)$ is in the order of a few GHz. We will now make two assumptions in order to clarify our analysis: first, we will assume that the envelope $\Phi(\Omega)$ can be considered approximately constant within our photodetection bandwidth. Secondly, we assume that all our measurements involve RF frequencies much below the repetition rate of the laser, which allows us to neglect all the harmonics in equation (\ref{input_signal})  except for the zero-frequency term. These assumptions allow us to simplify Eq. (\ref{input_signal}) as
	\begin{equation} \label{input_signal_simplified}
	\mathcal{S}_{\rm in}(\Omega) \approx P_0^2 \, \delta(\Omega),
	\end{equation}
	where $P_0=\langle \hat{N}_{\rm in}(0)\rangle$ is the mean (zero-frequency) photon-flux.
	
	To compute the noise term, $\mathcal{N}_{\rm in}(\Omega)$, we would require the auto-correlation function of the input photon flux, $r_{\rm in}(t, t') = \langle \hat{n}_{\rm in}(t) \, \hat{n}_{\rm in}(t') \rangle$, which we have not provided yet. This function is not readily accesible experimentally, and thus it is more convenient to postulate a noise spectrum directly in the frequency regime, from which the corresponding correlation function could be computed. In our experiment, we model the input probe noise as
	\begin{equation}\label{input_noise}
	\mathcal{N}_{\rm in}(\Omega)  = \mathcal{N}_{\rm in}^q + \mathcal{N}_{\rm in}^t(\Omega),
	\end{equation}
	where we assumed a frequency-constant component, $ \mathcal{N}_{\rm in}^q$, which describes the quantum fluctuations (shot-noise) of the probe  \cite{Bachor98}, together with an additive low-frequency component, $\mathcal{N}_{\rm in}^t(\Omega)$, which describes technical noise \cite{Ivanov03}. For the purposes of this discussion, we will just assume that the quantum component of the noise has a power spectrum proportional to the average photon flux, 
	\begin{equation}\label{input_noise_quantum}
	\mathcal{N}_{\rm in}^q = \kappa F P_0
	\end{equation}
	where $\kappa$ is a proportionality constant that accounts for the particularities of the experiment  (e.g. spectrum of the mode-locked laser). In addition, we have included a noise factor $F$, which is equal to 1 for Poissonian light, and takes higher and lower values for super and sub-Poissonian light respectively. The technical noise component, $\mathcal{N}_{\rm in}^t(\Omega)$, describes the low-frequency (< 1MHz) intensity fluctuations of classical origin that typically exist in a mode-locked laser \cite{Ivanov03}. 
	
	Let us now compute the signal and noise for the probe \textit{after} the SEM process. Neglecting higher harmonics, we write the modulated gain as
	\begin{equation}\label{gain_modulation}
	G_{\rm SE}(t) = (1+\Delta) + \Delta \, \cos(\Omega_0 t)
	\end{equation}
	where $\Delta = (G_0 - 1)/2$, with $G_0$ the maximum attained gain. The RF modulation frequency, $\Omega_0$, is within the bandwidth of our photodiode, while higher than the bandwidth of any technical noise in the experiment. Given this time-varying gain, the input and output frequency-resolved photon fluxes are simply related through
	\begin{equation}\label{N_out}
	\hat{N}_{\rm out}(\Omega) = (1+\Delta) \, \hat{N}_{\rm in}(\Omega) + \frac{\Delta}{2} \Big( \hat{N}_{\rm in}(\Omega - \Omega_0) + \hat{N}_{\rm in}(\Omega + \Omega_0) \Big),
	\end{equation}
	where the RF side-bands generated by the modulated gain is apparent. 
	
	We compute the \textit{signal} component of the PSD of the photocurrent applying (\ref{PSD_signal_component}) on (\ref{N_out}), which yields
	\begin{equation}
	\mathcal{S}_{\rm out}(\Omega) = (1+\Delta)^2 \, P_0^2 \, \delta(\Omega) + \frac{\Delta^2}{4} \ P_0^2 \ \delta(\Omega-\Omega_0)
	\end{equation}
	where it becomes clear that we can infer the stimulated emission gain by measuring the narrowband peak that appears at modulation frequency $\Omega_0$. Our analysis will consider only the content of the PSD around $\Omega_0$, which allows us to neglect the zero-frequency component and write
	\begin{equation}\label{modulated_signal}
	\mathcal{S}_{\rm out}(\Omega) =  (G_0 -1)^2 \ P_0^2 \ \delta(\Omega-\Omega_0)/16
	\end{equation}
	for our frequency range of interest.
	
	We now compute the \textit{noise} component of the output PSD by applying (\ref{PSD_noise_component}) to (\ref{N_out}). For frequencies $\Omega$ around $\Omega_0$, i.e not accounting for the PSD at baseband frequencies, we find
	
	\begin{subequations}\label{modulated_noise}
		\begin{align}
		&\mathcal{N}_{\rm out}(\Omega)  \\
		&= (1+\Delta)^2 \, \mathcal{N}_{\rm in}(\Omega) +  \frac{\Delta^2}{4} (\mathcal{N}_{\rm in}(\Omega-\Omega_0) +\mathcal{N}_{\rm in}(\Omega +  \Omega_0) ) \label{modulated_noise_1}\\
		& = \mathcal{N}_{\rm in}^q \ \Big((1+\Delta)^2 + \frac{\Delta^2}{2}\Big) + \frac{\Delta^2}{4} \mathcal{N}_{\rm in}^t(\Omega-\Omega_0) \label{modulated_noise_2} \\
		& \approx \mathcal{N}_{\rm in}^q \ (1+2\Delta) + \frac{\Delta^2}{4} \mathcal{N}_{\rm in}^t(\Omega-\Omega_0) \label{modulated_noise_3}  \\
		& = \mathcal{N}_{\rm out}^q+ \mathcal{N}_{\rm out}^t(\Omega)\label{modulated_noise_4}
		\end{align}
	\end{subequations}
	where in (\ref{modulated_noise_1}) we have assumed that different frequencies of the input probe are uncorrelated, such that $\langle \hat{N}_{\rm in}(\Omega) \hat{N}_{\rm in}(\Omega') \rangle = 0$ for $\Omega\neq \Omega'$. To arrive to (\ref{modulated_noise_2}), we have used equation (\ref{input_noise}) together with the assumption that $\mathcal{N}_{\rm in}^t(\Omega)\ne0$ only for $\Omega \ll \Omega_0$. To arrive to equation ((\ref{modulated_noise_3}) we have simply neglected the quadratic powers of $\Delta$ in the first term of the noise, since the modulation is assumed to be small. Finally, in (\ref{modulated_noise_4}) we have expressed the noise component of the PSD in terms of the output quantum fluctuations, $\mathcal{N}_{\rm out}^q = \kappa \, F P_0 \, G_0$, and the output technical noise, $\mathcal{N}_{\rm out}^t(\Omega) = (G_0 -1)^2 \mathcal{N}_{\rm in}^t(\Omega-\Omega_0)/16$.
	
	The signal-to-noise ratio (SNR) of the measurement can be computed using equations (\ref{modulated_signal}) and (\ref{modulated_noise_3}), from which we obtain:
	\begin{equation} \label{SNR_appendix}
	{\rm SNR} = \frac{\mathcal{S}_{\rm out}(\Omega_0) }{\mathcal{N}_{\rm out}(\Omega_0)} = \frac{(G_0-1)^2 P_0 /\kappa}{16 F G_0 + (G_0-1)^2 \rho_t}
	\end{equation}
	where we have defined $\rho_t = \mathcal{N}_{\rm in}^t(0) / (P_0 \kappa)$, which corresponds to the ratio between the spectral amplitude of the low-frequency technical noise and the quantum shot noise in a coherent state ($F=1$). This expression shows that the SNR of the SEM measurement can be efficiently improved by using sub-Poissonian light as a probe, as long as $16 F G_0 > (G_0-1)^2 \rho_t$. In conclusion, for small values of the gain (e.g. $G_0 < 1+10^{-2}$ in our experiment), the modulated technical noise can in practice be neglected in comparison with the wideband quantum shot noise, and the measurement can be effectively improved by using a sub-Poissonian probe.
	
	\section{Intensity squeezed light via single pass optical parametric amplification of coherent states }
	\label{sub_poissonian_statistics_derivation}
	In the Heisenberg picture, the single-mode parametric amplification of a coherent state is expressed by subsequently applying the displacement and squeezing operators on the vacuum state, resulting in the following evolved annihilation operator:
	\begin{equation}
	\label{single_mode_Heisenberg}
	\hat a_{r,\alpha} =  {\rm cosh}(r)(\hat a + \alpha) - {\rm sinh}(r) (\hat a^\dagger + \alpha^*),
	\end{equation} 
	where $\hat{a}$ is an annihilation operator acting on the vacuum state. Here $\alpha$ is the complex amplitude of the seed coherent state, and $r = \sqrt{P_{\rm p}} \, \beta $ is the squeezing parameter of the OPA, with $P_{\rm p}$ the pump power, and $\beta$ an effective nonlinearity.  The mean and variance of the photon number of the above squeezed coherent state, $\hat{n}_{r, \alpha} = \hat{a}_{r, \alpha}^\dagger \hat{a}_{r, \alpha}$, are 
	\begin{subequations} \label{appendix_n_var_transformations}
		\begin{align}
		\langle \hat n_{r, \alpha}\rangle &=& \Re[\alpha]^2 e^{-2r} + \Im[\alpha]^2 e^{2r} + {\rm sinh}^2(r), \label{nbar:SD} \\
		{\rm var}(\hat n_{r, \alpha}) &=& \Re[\alpha]^2 e^{-4r} + \Im[\alpha]^2 e^{4r} + \frac{1}{4}{\rm sinh}^2(2r),
		\end{align}
	\end{subequations} 
	where $\Re[\alpha]$ and $\Im[\alpha]$ represent real and imaginary parts of $\alpha$, respectively.  The average intensity after the OPA  is composed of the mean intensity of the input coherent seed, $|\alpha|^2$, amplified or deamplified depending on its phase, as well as of a stimulated emission contribution (last term in Eq. (\ref{nbar:SD}). The intensity fluctuations experience a similar behaviour, albeit with a different scaling on the squeezing coefficient -- which is responsible for sub-Poissonian intensity statistics. The intensity gain in the parametric amplification process, ${\cal G} \equiv \langle \hat{n}_{r, \alpha} \rangle / |\alpha|^2 $, is suscinctly written by expressing the input coherent amplitude as $\alpha = \lvert \alpha\rvert \, {\rm exp}(i\phi)$, and neglecting the spontaneous emission contribution:
	\begin{equation} \label{1mIG}
	{\cal G} \approx {\rm cos}^2(\phi)\, e^{-2r} + {\rm sin}^2(\phi)\, e^{2r}.
	\end{equation}   
	We quantify the intensity noise via the Fano factor $F$, which gives the ratio of the photon number variance to that expected for a coherent state, and is defined as \begin{equation} \label{1mff}
	F = \frac{{\rm var} (\hat n_{r, \alpha})}{\langle \hat n_{r, \alpha}\rangle} \approx \frac{{\rm cos}^2(\phi)\, e^{-4r} + {\rm sin}^2(\phi)\, e^{4r}}{{\rm cos}^2(\phi)\, e^{-2r} + {\rm sin}^2(\phi)\, e^{2r}}.
	\end{equation}
	By varying the phase of the input seed, $\phi$, we go from maximum intensity deamplification (${\cal G}<1$) and squeezing ($F < 1$) for $\phi = m\pi$ with $m \in \mathbf{Z}$ , to maximum intensity amplification (${\cal G}>1$) and antisqueezing ($F > 1$) for $\phi = m\pi + \pi/2$.
	
	In the single-mode model presented above, there is a one-to-one correspondence between classical intensity deamplification and squeezing of the intensity fluctuations. However, if the mode of the seed and a mode which is being deamplified (squeezed) are not identical, for example, if there is a spatial mismatch of the pump and seed beams, this correspondence breaks down. While a complete discussion of this phenomenon is outside of the scope of this paper, here we will consider a simple model of OPA where a fraction $\chi \in[0,1]$ of the intensity of the seed undergoes the transformations described in Eqs. (\ref{appendix_n_var_transformations}), while the rest of the seed remains unchanged. This simple model of OPA  yields the following maximum/minimum average power of the amplified/deamplified seed ~\cite{HirKuw05} : 
	\begin{equation} \label{classic_amp_appendix}
	P_{\rm s}^{\rm out} \approx   \Big(1-\chi + \chi \exp\big(\pm \beta \sqrt{P_{p}}\big)\Big)P_{\rm s}^{\rm in},
	\end{equation}
	where we have identified $P_{\rm s}^{\rm in} = |\alpha|^2$ and  $P_{\rm s}^{\rm out} = \langle \hat{n}_{\, r, \ \chi \alpha}\rangle + \langle \hat{n}_{\,0, \  (1 -\chi)\alpha}\rangle$.  Simultaneaously, this model predicts the following maximum/minimum intensity Fano factor for the amplified/deamplified seed:
	\begin{equation}\label{multimode_Fano_appendix}
	\begin{split}
	F &= \frac{{\rm var} (\hat{n}_{r, \, \chi \alpha}) +{\rm var} (\hat{n}_{0, \, (1-\chi) \alpha})}{\langle \hat{n}_{r, \, \chi \alpha} \rangle + \langle \hat{n}_{0, (1- \chi)\alpha} \rangle} \\
	&\approx \frac{1 - \chi + \chi \, \exp(\pm 4 \,\sqrt{P_{\rm p}} \, \beta)}{1 - \chi + \chi \, \exp(\pm 2 \, \sqrt{P_{\rm p}} \,  \beta)}.
	\end{split}
	\end{equation}
	
	The effect of linear loss can be modelled as a beam-splitter with intensity transmittance $\eta$, which compounds the propagation loss and inefficient detection. This transforms the mean photon number and Fano factor as \cite{Li95}
	\begin{equation} \label{loss}
	\begin{split}
	\langle \hat{n}_{r, \alpha, \chi} \rangle &\rightarrow  \eta \ \langle \hat{n}_{r, \alpha, \chi} \rangle \\
	F &\rightarrow 1- \eta + \eta \,  F
	\end{split}
	\end{equation}
	which yields a resulting output average power and Fano factor for the seed:
	\begin{subequations}
		\begin{align}
		P_{\rm s}^{\rm out} &\approx   \Big(1-\chi + \chi \exp\big(\pm \beta \sqrt{P_{p}}\big)\Big) \, \eta \, P_{\rm s}^{\rm in}, \label{appendix_output_seed_power}\\
		F &\approx 1 - \eta + \eta \, \frac{1 - \chi + \chi \, \exp(\pm 4 \,\sqrt{P_{\rm p}} \, \beta)}{1 - \chi + \chi \, \exp(\pm 2 \, \sqrt{P_{\rm p}} \,  \beta)}. \label{appendix_output_fano_factor}
		\end{align}
	\end{subequations}

\end{document}